\documentclass[twocolumn, amsmath, amsfonts, amssymb, trackchanges]{aastex62}
\usepackage{mathtools}
\usepackage{natbib}
\usepackage{bm}

\newcommand{\DivU}{\ensuremath{\nabla\cdot\bm{u}}}

\newcommand{\grad}{\ensuremath{\nabla}}

\newcommand{\lilstressT}{\ensuremath{\bm{\bar{\bar{\sigma}}}}}

\newcommand{\approptoinn}[2]{\mathrel{\vcenter{
	\offinterlineskip\halign{\hfil$##$\cr
	#1\propto\cr\noalign{\kern2pt}#1\sim\cr\noalign{\kern-2pt}}}}}

\usepackage{color}

\graphicspath{{./}{figs/}{../tex/figs/}}

\received{2019 June 6}
\revised{2019 July 25}
\accepted{2019 July 26}
\submitjournal{ApJ}

\shorttitle{Entropy Rain: Thermals in Stratified Domains}
\shortauthors{Anders, Lecoanet, and Brown}

\defcitealias{lecoanet&jeevanjee2018}{LJ19}
\newcommand{\LJ}{\citetalias{lecoanet&jeevanjee2018}}

\begin{document}
\title{Entropy Rain: Dilution and Compression of Thermals in Stratified Domains}

\correspondingauthor{Evan H. Anders}
\email{evan.anders@colorado.edu}

\author[0000-0002-3433-4733]{Evan H. Anders}
\affil{Dept. Astrophysical \& Planetary Sciences, University of Colorado -- Boulder, Boulder, CO 80309, USA}
\affil{Laboratory for Atmospheric and Space Physics, Boulder, CO 80303, USA}
\author[0000-0002-7635-9728]{Daniel Lecoanet}
\affil{Princeton Center for Theoretical Science, Princeton, NJ 08544, USA}
\affil{Department of Astrophysical Sciences, Princeton, NJ 08544, USA}
\author[0000-0001-8935-219X]{Benjamin P. Brown}
\affil{Dept. Astrophysical \& Planetary Sciences, University of Colorado -- Boulder, Boulder, CO 80309, USA}
\affil{Laboratory for Atmospheric and Space Physics, Boulder, CO 80303, USA}

\begin{abstract}
Large-scale convective flows called giant cells were once thought to transport the Sun's luminosity in the solar convection zone, but recent observations have called their existence into question.
In place of large-scale flows, some authors have suggested the solar luminosity may instead be transported by small droplets of rapidly falling, low entropy fluid.
This ``entropy rain'' could propagate as dense vortex rings, analogous to rising buoyant thermals in the Earth's atmosphere.
In this work, we develop an analytical theory describing the evolution of dense, negatively buoyant thermals.
We verify the theory with 2D cylindrical and 3D cartesian simulations of laminar, axisymmetric thermals in highly stratified atmospheres.
Our results show that dense thermals fall in two categories: a \emph{stalling} regime in which the droplets slow down and expand, and a \emph{falling} regime in which the droplets accelerate and shrink as they propagate downwards.
We estimate that solar downflows are in the falling regime and maintain their entropy perturbation against diffusion until they reach the base of the convection zone.
This suggests that entropy rain may be an effective nonlocal mechanism for transporting the solar luminosity.
\end{abstract}


\section{Introduction}
\label{sec:intro}
Recent observations of solar convection have revealed a convective conundrum.
Power spectra of horizontal velocities show weaker flows than anticipated at large length scales \citep{hanasoge&all2012, greer&all2015}.
These observations cast doubt on the existence of giant cells driven by deep convection which should manifest as powerful, large-scale motions at the solar surface. 
This discrepancy between theory and observations has called into question our fundamental understanding of convection, sparking numerous targeted investigations into the nature of solar convection  \citep{featherstone&hindman2016, omara&all2016, cossette&rast2016, kapyla&all2017, hotta2017}.

Rather than appealing to giant cells, \cite{spruit1997} hypothesized that convective motions in the Sun may be primarily driven by cooling in narrow downflow lanes at the solar surface.
\citet{brandenburg2016} incorporated this ``entropy rain'' concept into a non-local mixing length theory, and suggested the entropy rain could take the form of propagating vortex rings.
The entropy rain hypothesis assumes these small vortex rings can maintain their entropy perturbation as they traverse the entire solar convection zone.
This allows them to transport the solar luminosity via enthalpy fluxes.
However, the vortex rings described by \citet{brandenburg2016} do not include a fundamental aspect of entropy rain: entropy perturbations relative to the background atmosphere.
Entropy rain is dense, and buoyancy forces will modify its dynamics.

It is important to understand how the propagation of these basic convective elements is affected by their negative buoyancy. 
In the context of Earth's atmosphere, ``thermals,'' or buoyant fluid regions which evolve into rising vortex rings, are thought to be the basic unit of convection \citep[e.g.,][]{romps&all2015}. 
Atmospheric thermals are buoyant and rise, but the term is also used for dense, falling fluid.
We thus study the entropy rain hypothesis by investigating the evolution of individual dense thermals.

Thermals in the Boussinesq limit have been well studied in the laboratory for decades \citep[see e.g.][]{morton&all1956, scorer1957}, and more recently through Direct Numerical Simulation \citep[DNS,][]{lecoanet&jeevanjee2018}. 
These studies find that thermals expand radially and decelerate as they propagate.
Such a deceleration may cause the thermals to move so slowly they would diffuse away their entropy perturbation before reaching the bottom of the solar convection zone.
\citet{brandenburg&hazlehurst2001} found that hot, buoyant thermals in stratified domains behave qualitatively similar to Boussinesq thermals.
However, we are not aware of past work which carefully examines the effects of stratification on \emph{negatively} buoyant thermals.

\begin{figure}[t!]
    \includegraphics[width=\columnwidth]{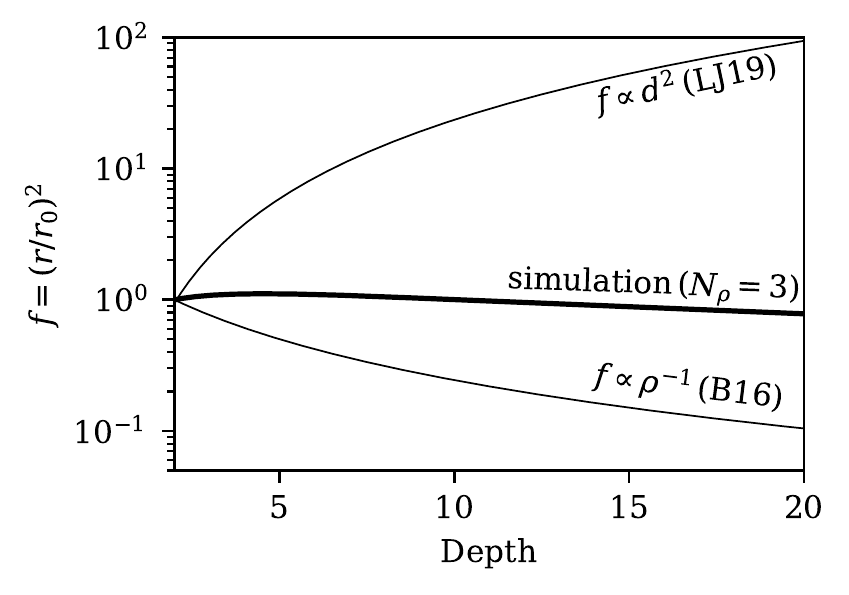}
    \caption{
	Shown is the filling fraction of a dense thermal as a function of depth in an atmosphere which spans three density scale heights (the $N_\rho = 3$ case examined later in this work). 
	Thin solid lines indicate the predictions for filling fraction growth in the Boussinesq case (\LJ) and for pure horizontal compression \citep[B16,][]{brandenburg2016}.
    \label{fig:overview} }
\end{figure}

Ignoring entropy variations, \citet{brandenburg2016} suggests the filling fraction $f$ of entropy rain should decrease like $f \propto \rho^{-1}$ for horizontal compression and $f \propto \rho^{-2/3}$ for spherical compression, where $\rho$ is the density.
On the other hand, the filling fraction of Boussinesq thermals \emph{increases} like $f \propto d^2$, where $d$ is the depth of the thermal.
These regimes are shown in Fig.~\ref{fig:overview} and compared to the true propagation of a numerically simulated dense thermal which includes both entropy variations and density stratification.

In this paper, we extend \citet{lecoanet&jeevanjee2018} (hereafter \LJ) to study the propagation of low-Mach number, low-entropy thermals in stratified domains. 
We are specifically interested in how the buoyancy force affects the scaling of the thermal radius, or filling fraction, with depth. 
If buoyancy dominates, it is possible that entropy rain would simply grow too large and stall before reaching the bottom of the solar convection zone.
On the other hand, if the compression effects of \citet{brandenburg2016} are dominant, then these thermals could propagate to the bottom of the solar convection zone, validating the entropy rain picture.

In section \ref{sec:theory}, we develop a theoretical description of thermals in a stratified domain. 
In section \ref{sec:experiment}, we describe the numerical experiments conducted in this work. 
In section \ref{sec:results}, we compare our theory and simulation results. 
In section \ref{sec:implications}, we discuss what our results imply for the entropy rain hypothesis.
Finally, in section \ref{sec:discussion}, we summarize our findings and conclusions.

\section{Model of thermal evolution}
\label{sec:theory}

\subsection{Phenomenological description of thermal evolution}
\begin{figure}[t!]
    \includegraphics[width=\columnwidth]{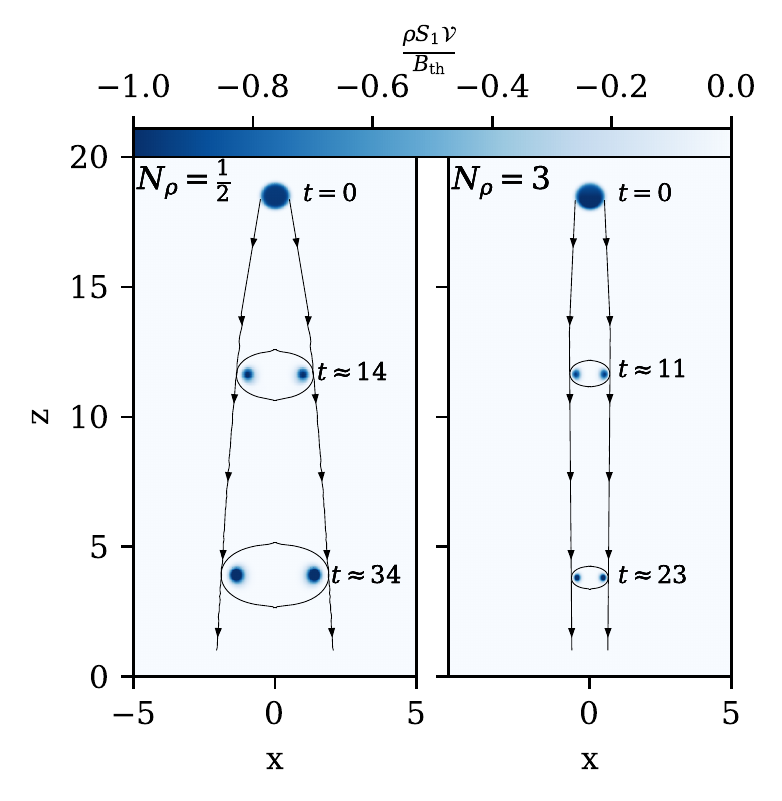}
    \caption{
	Shown is the evolution of entropy perturbations, normalized by the thermal's buoyancy perturbation, from 3D simulations conducted in this work.
	(left) A thermal in a weakly stratified domain with $N_\rho = 1/2$ density scale heights, and (right) a thermal in a strongly stratified domain with $N_\rho = 3$.
	While both start with precisely the same initial condition, the thermal in low stratification expands with depth and slows down, whereas the thermal in strong stratification compresses with depth and accelerates.
    \label{fig:evolution_colormeshes} }
\end{figure}

In Fig.~\ref{fig:evolution_colormeshes}, we show snapshots of 3D simulation data depicting the descent of cold thermals released from rest in two domains which span a different number of density scale heights, $N_\rho$.
The left panel shows a weakly stratified domain with $N_\rho=0.5$, whereas the right panel shows a strongly stratified domain with $N_\rho=3$.
In both cases, the thermal is initialized with a spherical negative entropy perturbation whose diameter is 5\% of the domain depth.
This dense sphere spins up into an axisymmetric vortex ring, and the vertical cross section through this vortex ring shows two circular vorticity and entropy extrema.
In the $N_\rho = 0.5$ simulation, the thermal grows with depth, similar to thermals in the Boussinesq regime.
On the other hand, in the $N_\rho = 3$ simulation, the thermal's radius decreases marginally with depth.

The goal of this paper is to understand the evolution of the thermal in the vortex ring stage.
All of the thermals studied in this work are laminar, similar to the Hill vortices studied by \citet{brandenburg2016}.
Crucially, \LJ\, showed little difference between the evolution of laminar and turbulent thermals in the Boussinesq limit.
As a result, we leave studies of turbulent thermals in stratified domains for future work.

In the following sections, we will use the impulse of dense vortex rings to derive expressions for the evolution of their depth and radii with time.
In this work we study vortex rings generated by discrete cold thermals, but ``plumes'' driven by time-stationary cooling produce similar vortex ring structures \citep[as in e.g.,][]{rast1998}.
The following description of vortex ring evolution should therefore be broadly applicable.

\subsection{Impulse}

The evolution of thermals in the Boussinesq limit has been understood for decades (see e.g.,~\LJ\, for a description and references).
While many theoretical descriptions rely on self-similarity arguments, we will show how the impulse of a thermal controls its evolution.
This section parallels a similar analysis for the Boussinesq case, presented in \citet{mckim&all2019}.

The hydrodynamic impulse is defined as \citep{shivamoggi2010},
\begin{equation}
\bm{I} = \frac{1}{2}\int_{\mathcal{V}} \bm{x}\times(\grad\times(\rho\bm{u}))dV,
\end{equation}
where $\bm{x}$ is the position vector and $\bm{u}$ is the fluid velocity. 
The impulse is the time-integrated work which has acted on the fluid to result in the current fluid motion.
It is thus unaffected by internal forces (e.g., pressure or viscous).
Upon integration by parts, it is obvious that the impulse encompasses the momentum of the fluid within the volume $\mathcal{V}$.
However, one can also show that the surface terms correspond to the momentum outside the volume $\mathcal{V}$ \citep[e.g.][]{Akhmetov2009}.
A thermal with volume $\mathcal{V}$, density $\rho$, and translating with velocity $\bm{u} = w_{\text{th}} \hat{z}$ thus has an impulse
\begin{equation}\label{eqn:momentum}
I_z = (1+k) \rho \mathcal{V} w_{\text{th}},
\end{equation}
where $k$ encompasses the ``virtual mass effect'' from the environmental fluid moving together with the thermal \citep{tarshish&all2018}.

We now restrict our study to an ideal gas in the low Mach-number regime, in an adiabatic background.
This is the appropriate regime of deep solar convection.
Due to rapid pressure equilibration in low Mach-number flows, we can approximate 
\begin{equation*}
\frac{\rho_1}{\rho_0} \approx -\frac{S_1}{c_P},
\end{equation*}
where $S_1$ is the specific entropy perturbation and $c_P$ is the specific heat at constant pressure; thermodynamic variables are decomposed into background (subscript 0) and fluctuating (subscript 1) components.

The rate of change of the impulse is
\begin{equation*}
\frac{d\bm{I}}{d t} = \int_\mathcal{V} \rho_1\, \bm{g}\, dV,
\end{equation*}
because the surface terms completely cancel.
Assuming a uniform, vertical gravity, $\bm{g} = -g\hat{z}$, it is useful to define the buoyancy perturbation,
\begin{equation}
B \equiv \int_{\mathcal{V}} \rho_0\, S_1\, \frac{g}{c_P}\, dV.
\label{eqn:tot_buoyancy}
\end{equation}
such that
\begin{equation}
\frac{d I_z}{d t} = B.
\label{eqn:change_in_impulse}
\end{equation}

In the limit of a low Mach-number, thin-core vortex ring, the impulse can be approximated as
\begin{equation}
I_z \approx \pi \rho_0 r^2 \Gamma,
\label{eqn:impulse_approx}
\end{equation}
where $r$ is the radius of the thermal from its axis of symmetry to its vorticity extremum, and $\Gamma = \int_{\mathcal{A}} (\grad\times\bm{u})\cdot d\bm{A}$ is the circulation in a cross-section of the vortex ring.
The circulation can change due to baroclinic torques,
\begin{equation}
\frac{d\Gamma}{dt} = \oint_{\mathcal{C}} g \frac{S_1}{c_P}\hat{z} \cdot d\bm{x},
\label{eqn:circulation}
\end{equation}
where $\mathcal{C}=\partial \mathcal{A}$ is the contour around the thermal's vorticity.
For the case of a vortex core in which the entropy perturbation is contained tightly in the core, as in Fig.~\ref{fig:evolution_colormeshes}, a contour can be drawn for which $S_1\approx0$.
Thus, there are no net baroclinic torques, and we will treat the circulation of a developed vortex ring as a conserved quantity.

\subsection{Model of thermal evolution}
\begin{figure}[t!]
    \includegraphics[width=\columnwidth]{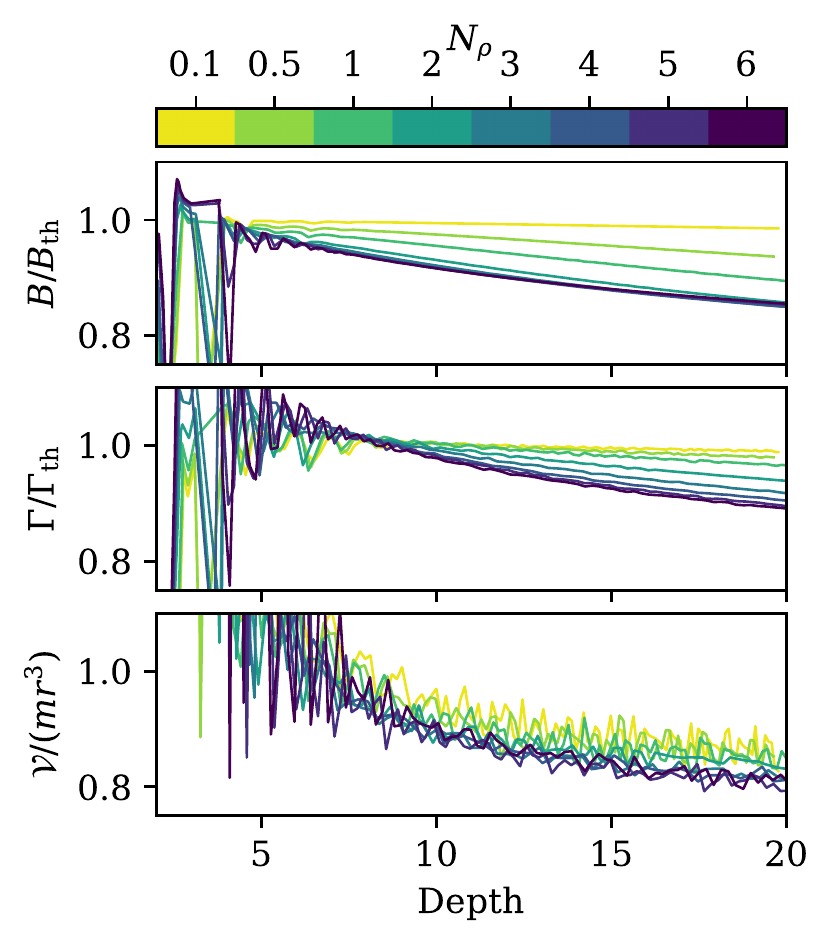}
    \caption{
	Shown is the evolution of buoyancy perturbation (top), circulation (middle), and volume factor (bottom) for each 2D simulation conducted in this work.
	All three of these quantities remain nearly constant after an initial spin-up phase.
	With increasing stratification, we see marginally more detrainment (loss of buoyant signature) in the top panel.
    \label{fig:constants} }
\end{figure}

The thermals studied here begin as initial spherical perturbations and spin up into vortex rings.
The spun up vortex ring phase can be modeled as having evolved from a ``virtual origin'' where the vortex ring had zero radius.
We model the thermal as having been located at its virtual origin at a temporal offset $t = -t_{\text{off}}$, where $t = 0$ is the time at which the true thermal was released from rest.

In our simulations, we find the thermal undergoes weak detrainment (loss of buoyant signature to environmental fluid), so the negative buoyancy of the thermal decreases slightly in time (Fig.~\ref{fig:constants}, top panel).
We thus express the buoyancy perturbation as 
\begin{equation}
B \approx \eta B_{\text{th}},
\end{equation}
where $\eta$ is a constant of $\mathcal{O}$(1) which represents this detrainment and $B_{\text{th}}$ is the thermal's characteristic buoyancy perturbation. 
We then integrate Eqn.~\ref{eqn:change_in_impulse},
\begin{equation*}
I_z = \eta B_{\text{th}} (t + t_{\text{off}}).
\end{equation*}

Combining this with Eqn.~\ref{eqn:impulse_approx}, we retrieve our first main result,
\begin{equation}
r = \sqrt{\frac{\eta B_{\text{th}} (t + t_{\text{off}})}{\pi\rho_0\Gamma_{\text{th}}}},
\label{eqn:r_theory}
\end{equation}
Here, $\Gamma_{\text{th}}$ is the characteristic circulation of the thermal. As there are no net baroclinic torques, $\Gamma_{\text{th}}$ remains nearly constant in our simulations (Fig.~\ref{fig:constants}, middle panel).
In the Boussinesq limit where $\rho \rightarrow \text{constant}$, we retrieve the $r \propto \sqrt{t}$ scaling found in \LJ.
In the limit of strong stratification, we find $r \propto \rho^{-1/2}$, corresponding to purely horizontal compression, with $r^2 \propto \rho^{-1}$ \citep{brandenburg2016}.

To solve for the vertical evolution of the thermal, we can use Eqn.~\ref{eqn:momentum}.
We approximate the volume as $\mathcal{V} \approx m r^3$, where $m$ is a parameter which we take to be constant (which is a decent assumption in our simulations; see Fig.~\ref{fig:constants}, bottom panel).
Here, $m$ accounts for volumetric constants (e.g., $4\pi/3$), the aspect ratio of the thermal, and the cubed ratio between the full radius of the spheroidal thermal and $r$.
Defining the thermal velocity $w_{\text{th}} \equiv dz_{\text{th}}/dt$, we find
\begin{equation}
\frac{dz_{\text{th}}}{\rho(z_{\text{th}})^{1/2}} =
\left(\frac{(\pi\Gamma_{\text{th}})^{3/2}}{m(1 + k)(\eta B_{\text{th}})^{1/2}}\right)\frac{dt}{(t + t_{\text{off}})^{1/2}}.
\label{eqn:dz_theory}
\end{equation}
This can be easily integrated given an atmospheric stratification $\rho(z)$.

To summarize, we model thermals as thin-core vortex rings.
The vortex ring is parameterized by its buoyancy perturbation and circulation, which are assumed to be nearly constant after spin-up.
The impulse increases in magnitude due to buoyancy forces (Eqn.~\ref{eqn:change_in_impulse}), and allows us to relate the size of the vortex ring (Eqn.~\ref{eqn:impulse_approx}) to the momentum of the thermal and its ambient fluid (Eqn.~\ref{eqn:momentum}).
Assuming the thermal volume is spheroidal and that the virtual mass effect and detrainment can be parameterized as constants, we arrive at Eqn.~\ref{eqn:dz_theory}.

\subsection{Polytropic atmosphere solution}
In this work, we study an ideal gas with an adiabatic index of $\gamma = 5/3$.
An adiabatic polytrope satisfies
\begin{gather}\label{eqn:T0}
T_0 = 1 + (z - L_z)\grad_{\text{ad}}, \\
\rho_0 = T_0^{\,n_{\text{ad}}},
\label{eqn:polytrope}
\end{gather}
where $n_{\text{ad}} = (\gamma-1)^{-1}$ and $\grad_{\text{ad}}$ is the adiabatic temperature gradient.
All thermodynamic quantities are nondimensionalized such that $P_0 = \rho_0 = T_0 = 1$ at $z = L_z$, the top of the domain, where $T_0$ is the atmospheric temperature and $P_0 = \rho_0 T_0$ is the atmospheric pressure.

Integrating Eqn.~\ref{eqn:dz_theory} under this polytropic density stratification, we find
\begin{equation}
z_{\text{th}} = \grad_{\text{ad}}^{-1}\left[\left(\frac{2C}{ \alpha } \sqrt{t + t_{\text{off}}} + T_{\text{th},0}^{1/\alpha}  \right)^{\alpha} - 1\right] + L_z,
\label{eqn:theory_z}
\end{equation}
where 
$$
C \equiv \frac{\grad_{\text{ad}}\Gamma_{\text{th}}}{m(1 + k)} \sqrt{\frac{\pi^3\Gamma_{\text{th}}}{\eta B_{\text{th}}}}.
$$ 
The thermal is at the virtual origin, $z=z_{\text{th, 0}}$, at time $t=-t_{\text{off}}$, and the temperature there is $T_{\text{th},0} = 1 + (z_{\text{th},0} - L_z)\grad_{\text{ad}}$.
We define $\alpha^{-1} \equiv 1 - n_{\text{ad}}/2$, and in the limit of large stratification, we find that $z_{\text{th}} \propto t^2$ for our case of $\alpha = 4$. 

In our simulations, the thermal is initialized as a uniform sphere of dense fluid but it quickly spins up into a vortex ring. 
While we do not attempt to model the spin-up phase in this paper, it can be parameterized by the buoyancy $B_{\text{th}}$, circulation $\Gamma_{\text{th}}$, as well as the virtual origin $z_{\text{th,0}}$, and temporal offset $t_{\text{off}}$. 
Our theory also involves the volumetric aspect ratio of the thermal, $m$, the detrainment fraction, $\eta$, and the effective buoyancy, $k$. 
These appear to be only weakly dependent on the stratification for the thermals we have simulated.

\section{Simulation setup} 
\label{sec:experiment}
To test our theory, we run a series of thermal simulations using the 3D fully compressible equations in cartesian domains.
We additionally compute 2D azimuthally-symmetric simulations using the anelastic equations in cylindrical domains.
We verify the 3D and 2D simulations produce the same results when run with the same parameters.
Because 2D simulations are less computationally expensive, we use them to cover a broader parameter regime.

While solar convection is very turbulent, we restrict our study to laminar thermals. 
In the Boussinesq limit, \LJ\, showed the evolution of turbulent vortex rings is well described by laminar theory; we expect this will also hold in the stratified case.
In future work, we will apply this laminar theory to turbulent thermals with density stratification, which necessitates 3D simulations.

\subsection{2D Anelastic Simulations}
The LBR anelastic equations are \citep{lecoanet&all2014},
\begin{gather}
\DivU = -w \partial_z \ln\rho_0, \\
\begin{split}
\partial_t \bm{u} + \bm{u}\cdot\grad&\bm{u} = \\
- \grad \varpi + S_1\hat{z} &
+ \frac{1}{\rho_0\text{Re}}\left[\grad^2 \bm{u} + \frac{1}{3}\grad(\DivU)\right] 
\end{split}\\
\begin{split}
\partial_t S_1 + \bm{u}\cdot\grad S_1 =& \\
\frac{1}{\text{Re}}\left(\frac{1}{\rho_0c_P\text{Pr} }\right.&[\grad^2 S_1 + \partial_z\ln T_0 \cdot\partial_z S_1]\\
&+ \left.\frac{-\grad_{\text{ad}}}{\rho_0 T_0}\sigma_{ij}\partial_{x_i}u_j \right),
\end{split}
\end{gather}
where $\lilstressT$ is the viscous stress tensor in units of inverse time.
We solve these equations in cylindrical geometry, assuming axisymmetry.

Following \LJ\,, we non-dimensionalize the equations on the initial diameter of the thermal and its freefall velocity.
These equations are then fully specified in terms of the Reynolds number and Prandtl number,
\begin{equation}
\text{Re} = \frac{ u_{\text{th}} L_{\text{th}}}{\nu}, \qquad
\text{Pr} = \frac{ u_{\text{th}} L_{\text{th}}}{\chi}, \qquad
u_{\text{th}}^2 = \frac{g L_{\text{th}} \Delta s}{c_P},
\end{equation}
where $u_{\text{th}}$ is the freefall velocity, $L_{\text{th}}$ is the thermal length scale, and
$\Delta s$ is the magnitude of the specific entropy signature of the thermal.

The background density and temperature are given by Eqs.~\ref{eqn:T0} \& \ref{eqn:polytrope}.
The adiabatic temperature gradient is $\nabla_{\text{ad}}= -g(e^{N_\rho/n_{\text{ad}}}-1)/(L_z c_P)$, where $L_z=20$ is the height of the domain, $g = c_P$ is the gravity, and $N_\rho$ is the number of density scale heights spanned by the domain.

We choose an atmospheric model in which the dynamic viscosity, $\mu = \rho_0 \nu$, and the thermal conductivity, $\kappa = \rho_0 \chi$, are both uniform in space and constant in time.
We make this choice because $\mu$ and $\kappa$ appear in the density-weighted momentum and entropy equations, and we find the density-weighted entropy and momentum to be key quantities in our thermal theory.
The diffusivities $\nu$ and $\chi$ therefore scale inversely with the density.
As the diffusivities scale with depth, Re is specified at the thermal's initial depth.
All simulations conducted in this work use an initial value of Re = 600 and $\mbox{\text{Pr} = 1}$.

\subsection{3D Fully Compressible Simulations}
In order to verify our 2D anelastic simulations, we also simulate thermals with the 3D Navier Stokes equations. 
We use the $(T, \ln\rho)$ formulation of the equations \citep{lecoanet&all2014, anders&brown2017},
\begin{gather}
\partial_t \ln\rho_1 + \epsilon^{-1}\left(\bm{u}\cdot\grad\ln\rho_0 + \DivU\right) = -\bm{u}\cdot\grad\ln\rho_1, \\
\begin{split}
\partial_t \bm{u}  + \bm{u}\cdot\grad\bm{u} + \frac{1}{-\grad_\text{ad}}[\grad T_1 + &T_1\grad\ln\rho_0 + T_0\grad\ln\rho_1]  =\\
- \frac{\epsilon}{-\grad_\text{ad}} T_1\grad\ln\rho_1 + \frac{1}{\rho\text{Re}}&\left[\grad^2\bm{u} + \frac{1}{3}\grad(\DivU)\right]
\end{split} \\
\begin{split}
&\partial_t T_1 + \epsilon^{-1}\left[\bm{u}\cdot\grad T_0 + (\gamma-1)T_0\DivU\right] = \\
&-\left[\bm{u}\cdot\grad T_1 + (\gamma-1)T_1\DivU\right] + \frac{-\grad_{\text{ad}}}{\rho c_V\text{Re}}\left[\frac{1}{\text{Pr}}\grad^2 T_1 + \sigma_{ij}\partial_{x_i}u_j\right].
\end{split}
\end{gather}
These equations are non-dimensionalized in the same way as the anelastic equations, and use the same background atmosphere.
The new parameter $\epsilon = -u_{\text{th}}^2/\grad_{\text{ad}}$ is the magnitude of entropy perturbations and sets the Mach number of the thermal flows analagously to the adiabatic excess in polytropic convection \citep{anders&brown2017}; we use $\epsilon = 10^{-4}$ in this work. 

\subsection{Initial conditions}
The simulations are initialized with a spherical specific entropy perturbation,
\begin{equation}
S_1 = \frac{1}{2}\left[\text{erf}\left(\frac{r' - r_{\text{th}}}{\delta}\right) - 1\right].
\label{eqn:thermal_IC}
\end{equation}
Here, $r' = \sqrt{r^2 + (z - z_0)^2}$, where $z_0 = L_z - 3r_{\text{th}}$, with the thermal radius set as $r_{\text{th}} = 0.5$, and a smoothing width, $\delta = 0.1$.
As mentioned previously, Re = 600 and Pr = 1 are specified at the thermal's initial depth, $z = z_0$.

For the fully compressible simulations, we must also specify the density perturbation $\rho_1$.
We pick perturbations that are in pressure equilibrium,
\begin{equation}
\ln\rho_1 = S_1/c_P, \qquad T_1 = T_0(e^{-\epsilon\ln\rho_1} - 1)/\epsilon.
\end{equation}
In all cases, we do not add any symmetry breaking perturbations (e.g., noise).

\subsection{Numerics}
We simulate the thermals using the  Dedalus\footnote{\url{http://dedalus-project.org/}} pseudospectral framework \citep{burns&all2016, burns&all2019}.
The 2D simulations use an implicit-explicit (IMEX), third-order, four-stage Runge-Kutta timestepping scheme RK443 \citep{ascher&all1997}, and the 3D simulations use the second-order semi-implicit backward differentiation formulation SBDF2 \citep{wang&ruuth2008}.

The 2D simulation domain is periodic in the z-direction with $z \in [-L_z/4, L_z]$ and the radial direction spans $r \in [0, L_r]$.
The boundary conditions are $\partial_r S_1 = w = (\grad\times\bm{u})_\phi = 0$ at $r = L_r$, and the regularity of the equations automatically impose $u = \partial_r(w) = \partial_r(S_1) = 0$ at $r = 0$.
The 3D simulation domain is periodic in the horizontal directions ($x, y \in [-L_r, L_r]$) and vertically spans $z \in [0, L_z]$.
We impose impenetrable, stress free, fixed-temperature boundary conditions at the upper and lower boundaries.
In all of our simulations we specify $L_z = 20$ and $L_r = 5$.
We extend our 2D simulation domains to $z = -L_z/4$ because those simulations are vertically periodic.
This extension allows us to study the full transit of the thermal above $z = 0$ and terminate the simulation before it begins to wrap through the bottom of the periodic domain.

In our 2D simulations, we represent the radial direction with Chebyshev polynomials so that we can include geometric factors in our equations and capture the singularity at $r = 0$.
We then assume the z direction is periodic to make the calculations more efficient, and we end the simulations before the thermal can interact with the vertical periodic boundaries.
Remarkably, the results of these simulations vary only minimally from our 3D simulations, as we will show in the next section, indicating that the vertical periodicity and the different side boundaries (cylinder vs. cube) do not influcence the thermal properties.

All of the code used to perform the simulations in this work can be found online in the supplementary materials in a Zenodo repository \citep{supp_andersetall2019} at \dataset[10.5281/zenodo.3311894]{https://doi.org/10.5281/zenodo.3311894}.

\section{Model verification}
\label{sec:results}
\begin{figure}[t!]
    \includegraphics[width=\columnwidth]{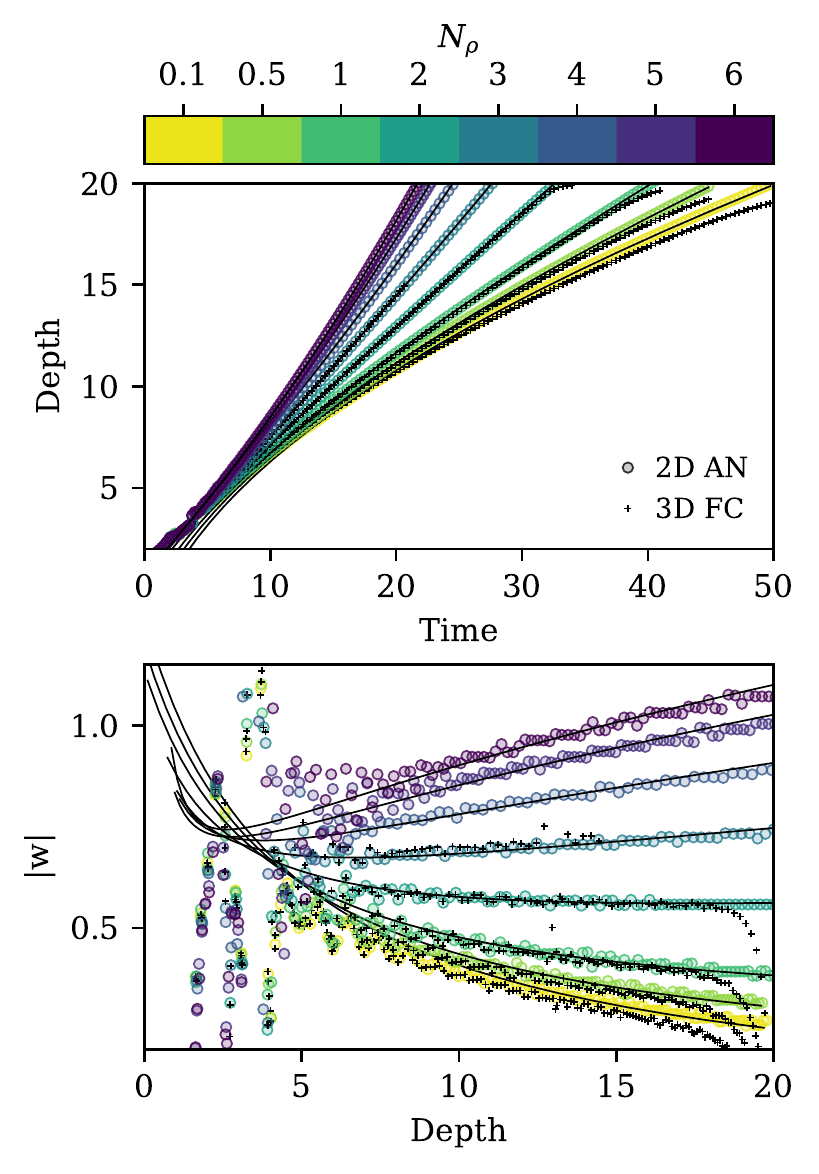}
    \caption{
	(top) The thermal depth as a function of time for the 2D anelastic (AN) and 3D fully compressible (FC) simulations.
	(bottom) The corresponding thermal velocities as a function of depth.
	Theoretical predictions from section \ref{sec:theory} are plotted in thin solid lines for each case (see parameters in table \ref{table:parameters}).
    \label{fig:results_panels} }
\end{figure}

\begin{figure}[t!]
    \includegraphics[width=\columnwidth]{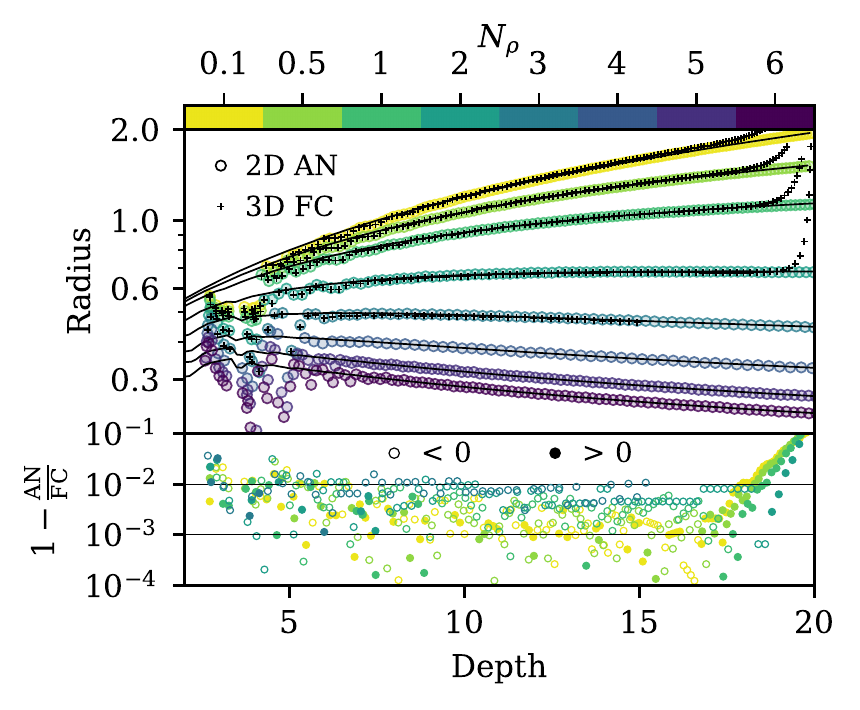}
    \caption{
	(top) Thermal radius as a function of depth for the 2D anelastic (AN) and 3D fully compressible (FC) simulations.
	Theoretical predictions from section \ref{sec:theory} are plotted in thin solid lines.
	(bottom) The fractional difference between anelastic and fully compressible simulations.
    \label{fig:diff} }
\end{figure}

To compare to the model, we must measure the thermal's depth and radius.
We define the thermal's depth and radius using the thermal's entropy minimum.
For specifics on how these measurements are conducted in our simulations, we refer the reader to appendix \ref{appendix:measurements}.

In the top panel of Fig. \ref{fig:results_panels} we show the depth, $d_{\text{th}} = L_z - z_{\text{th}}$, of the thermal as a function of time for simulations with different stratifications. 
At very low stratification (e.g., $N_\rho = 0.1$), the thermal is small compared to the local density scale height at all depths, and it evolves roughly according to the Boussinesq prediction of $d \propto \sqrt{t}$. 
As the stratification increases, the thermal transits the domain more quickly and approaches the limit of $d \propto t^2$ predicted in the highly stratified limit of Eqn.~\ref{eqn:theory_z}. 
The theoretical fits for depth from the prediction of Eqn.~\ref{eqn:theory_z} are plotted over the measured data.
The theoretical fits are poor at early times when the thermal is spinning up from its initial spherical state into the vortex ring state.
Once the thermal is spun up into a vortex ring, the theory shows remarkable agreement with the data.

In the bottom panel of Fig.~\ref{fig:results_panels} we show the corresponding thermal velocity as a function of depth.
Low density ($N_\rho = 0.1$) thermals decelerate with depth.
With increasing stratification, this deceleration stops and sufficiently stratified runs ($N_\rho \geq 3$) experience acceleration.

The top panel of Fig.~\ref{fig:diff} plots the thermal radius as a function of depth.
In the low stratification limit, the radius of the thermal grows linearly with depth, $r \propto d$, as is the case in the Boussinesq limit (\LJ).
The growth of the thermal is due to entrainment of environmental fluid, which causes the thermal to decelerate like $w \propto d^{-1}$, as is shown in the bottom panel of Fig.~\ref{fig:results_panels}.
However, as stratification increases, the thermal entrainment of environmental fluid decreases and it experiences less expansion.
In the limit of large stratification ($N_\rho \geq 3$), thermals contract with depth, and the thermals accelerate as they fall.

Finally, we quantify the excellent agreement between the 2D anelastic and 3D fully compressible simulations in Fig.~\ref{fig:diff}, bottom panel.
We ran simulations in both models for $N_\rho = [0.1, 0.5, 1, 2, 3]$.
There are slight discrepancies early in the simulation, when the thermal is spinning up, and late in the simulation, when the 3D simulations begin interacting with the bottom of the domain.
Outside these times, we find differences in the measured radius of less than 1\%.
This gives us confidence that our high stratification anelastic simulations are producing reliable results.
\citet{lecoanet&all2014} also found close agreement between low Mach-number anelastic and fully compressible simulations.

\begin{deluxetable*}{c c c c c c c c}
\tabletypesize{\footnotesize}
\caption{Simulation output parameterization
\label{table:parameters}
}
\tablehead{																																															
\colhead{$N_\rho$} & \colhead{$z_{\text{th},0}$} & \colhead{$t_{\text{off}}$} & \colhead{$B_{\text{th}}$} & \colhead{$\Gamma_{\text{th}}$} & \colhead{$m$} & \colhead{$\eta$} & \colhead{$k$}	}	
\startdata																																															
\multicolumn{8}{l}{2D Anelastic Simulations (Cylindrical)}\\
0.1 	&  24.6 	& 0.144	& -0.548 & -2.17 & 8.05	& 1.04	& 0.732	\\
0.5 	&  24.2 	& 0.695	& -0.569 & -2.12 & 8.34	& 0.977	& 0.715	\\
1	 	&  23.7 	& 1.11 	& -0.602 & -2.05 & 8.65	& 0.915	& 0.703	\\
2	 	&  22.3 	& 1.27	& -0.713 & -1.89 & 9.23	& 0.842 & 0.682	\\
3	 	&  21.2 	& 1.01	& -0.947 & -1.73 & 9.81	& 0.807	& 0.654	\\
4	 	&  20.5 	& 0.615	& -1.47	 & -1.59 & 10.2	& 0.794	& 0.642	\\
5	 	&  20.0	    & 0.425	& -2.70	 & -1.49 & 10.7	& 0.781	& 0.609	\\
6	 	&  19.8 	& 0.041	& -5.73	 & -1.43 & 10.8	& 0.787	& 0.616	\\
\multicolumn{8}{l}{3D Fully Compressible Simulations (Cartesian)}\\    
0.1 	&  23.4 	& -0.337& -0.547 & -2.17 & 8.98	& 1.06	& 0.636	\\
0.5 	&  23.8 	& 0.572	& -0.568 & -2.12 & 8.79	& 0.978	& 0.678	\\
1	 	&  23.6 	& 1.15	& -0.601 & -2.05 & 8.87	& 0.907	& 0.689	\\
2	 	&  22.4 	& 1.38	& -0.711 & -1.89 & 9.31	& 0.828	& 0.680	\\
3	 	&  21.1 	& 0.78	& -0.949 & -1.75 & 9.99	& 0.815	& 0.648	\\
\enddata																																															
\tablecomments{ 
Parameters presented in this table are best fits to simulation output data in the range $z = [0.1 L_z, 0.65 L_z]$.
Above this range, the thermal is still spinning up from initial conditions; below this range, 3D cases are heavily interacting with the bottom boundary.}
\end{deluxetable*}

The values of the parameters described in section \ref{sec:theory} for each of our simulations are presented in table \ref{table:parameters}.
As anticipated, $\eta$ is $\mathcal{O}$(1) and decreases slightly in value with increasing stratification, consistent with Fig.~\ref{fig:constants}.
In all cases, the buoyancy $B_{\text{th}}$ is similar to the integrated buoyancy in the initial conditions, with some losses due to detrainment in the spin-up.
We also find that the non-dimensional circulation is roughly $-2$ for each of our cases, and decreases in magnitude with increasing stratification.

\section{Implications for entropy rain hypothesis}
\label{sec:implications}
Our theory shows that the evolution of dense thermals in stratified domains is complex.
Neither the assumption of pure compression \citep[as in e.g.,][]{brandenburg2016} nor the evolution of thermals in the Boussinesq regime (\LJ) fully describes thermal behavior.
Rather, the results fall somewhere in between, and theory and simulations suggest that there are two regimes of downflowing thermal behavior:
\begin{enumerate}
\item A low-stratification ``stalling'' regime, in which the thermal entrains environmental fluid and slows down, acting much like the Boussinesq regime, and
\item A high-stratification ``falling'' regime, in which the thermal falls sufficiently fast that atmospheric compression dominates over entrainment and the thermal accelerates as it falls deeper into the atmosphere.
\end{enumerate}
We note that both of these regimes could have interesting implications for the entropy rain hypothesis.

If the solar downflows are in the stalling regime, convective elements would grow enormously in size and slow down very close to the solar surface.
In a perfectly quiescent atmosphere, these slow, large convective elements would eventually propagate to the base of the convection zone over long timescales.
In fact, their large length scales would likely help shield them from any dissipative effects despite their low velocities.
However, the solar convection zone is highly turbulent, and we expect that a more likely outcome for such large, coherent, and slowly propagating structures is that they would be torn apart by turbulent motions.

On the other hand, if solar convection is comprised of thermals in the falling regime, then it is likely that solar surface elements would reach deep into the Sun.
Neglecting buoyancy, we expect that downward propagating vortex rings in the solar convection zone would likely compress to sizes on which conductivity could become important.
Our theory of thermals suggests that, instead, buoyancy counteracts some of the compressional effects of stratification, and could help convective elements maintain their entropy perturbation as they cross the solar convection zone.

We now estimate the behavior of thermals in the Sun based on the simulations presented in this work.
The thermal evolution depends on a variety of parameters (see table \ref{table:parameters}).
However, we find the only parameter which changes appreciably as we increase the stratification is the normalized buoyancy $B_\text{th}$.
To estimate $B_\text{th}$ in the sun, we approximate the dimensional buoyancy perturbation of Eqn.~\ref{eqn:tot_buoyancy} as $\tilde{B} = \rho \mathcal{V} g (S_1/c_p)$, and calculate this quantity for thermals launched from the solar photosphere, assuming that solar downflow lanes quickly break up into thermals.
Using the VAL atmospheric data of \citet{avrett&loeser2008}, we estimate the solar surface to have a temperature of $T_0 \approx 6000$ K, a density of $\rho_0 \approx 1.74 \times 10^{-7}$ g/cm$^3$, and a sound speed of $c_s = 9.5 \times 10^5$ cm/s.
We estimate that thermal diameters would be roughly the width of downflow lanes ($L_{\text{th}} = 0.1$ Mm), and the average atmospheric density over the first $L_{\text{th}}$ of the solar interior is $\bar{\rho} = 1.17\rho_0$.
Estimating downflows to have a temperature deviation of $T_1 = -500$ K \citep{borrero&bellotrubio2002}, the nondimensional entropy signature of solar downflows is $S_1/c_p \sim \gamma^{-1}\ln(1 + T_1/T_0) = -5.22 \times 10^{-2}$.
Using a solar surface gravity value of $g = 2.74 \times 10^4$ cm/s$^2$ and assuming thermal formation occurs at the solar photosphere, the buoyancy of a spherical thermal is 
$$
\tilde{B} \approx \bar{\rho}\left[\frac{4\pi}{3} \left(\frac{L}{2}\right)^3\right] g \frac{S_1}{c_P} = -1.52 \times 10^{17} \text{g cm}^4/\text{s}^2.
$$
Nondimensionalizing this by $B_0 = (S_1/c_P) L^2 u_{\text{th}}^2 \rho_0$ with $u_{\text{th}} = c_s \sqrt{S_1/c_P}$, we find $B_\text{th} = \tilde{B}/B_0 = -3.57$, which lies between the $N_\rho = 5$ and $N_\rho = 6$ simulations we studied here (see table \ref{table:parameters}).

\begin{figure}[t!]
    \includegraphics[width=\columnwidth]{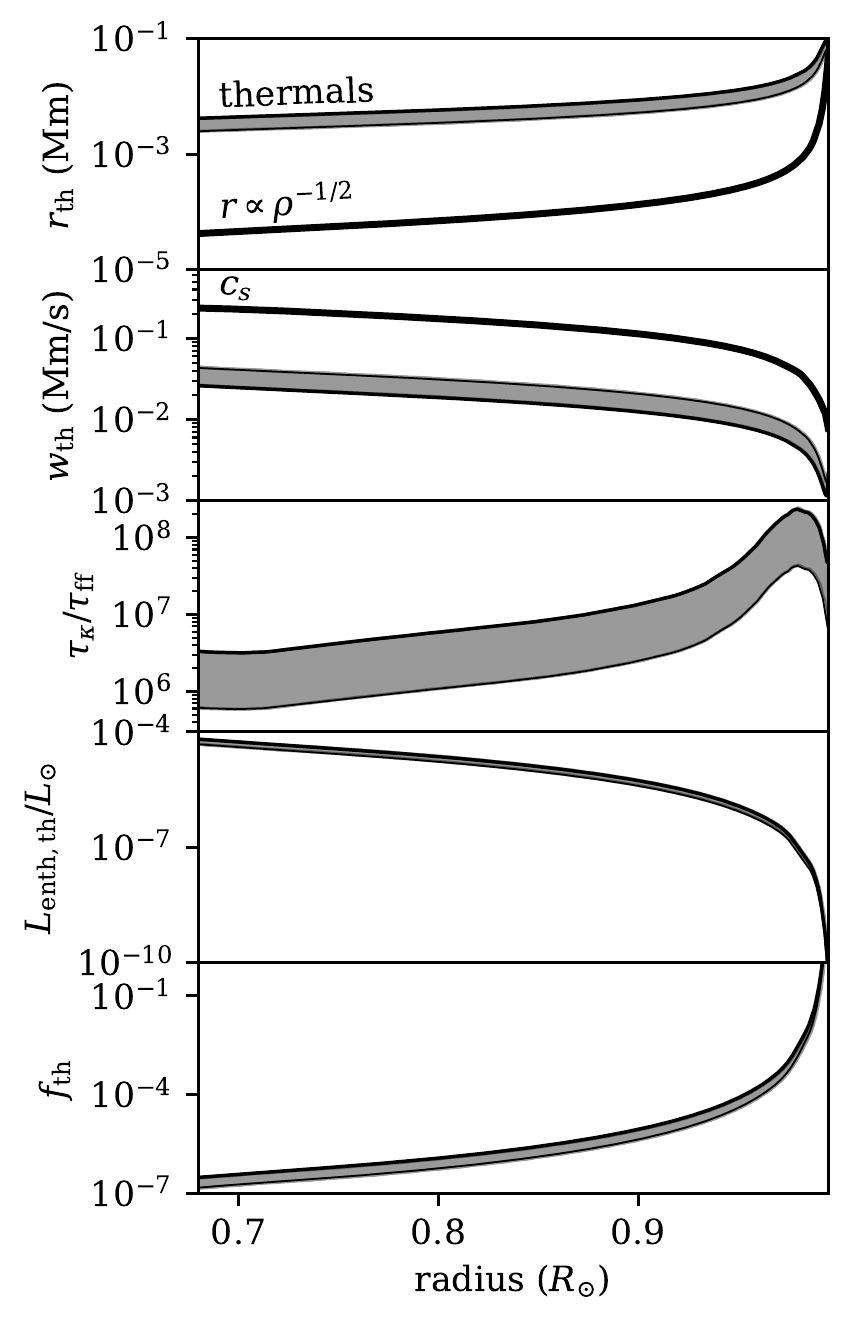}
    \caption{
	(First panel) Bounds on the evolution of the radii of entropy rain in the Sun are shown and compared to what the radii would be under purely horizontal compression.
	(Second panel) The corresponding range of velocities of solar-like thermals are shown. 
	(Third panel) Shown is the estimated range of the ratio of the thermal diffusion timescale normalized by the freefall timescale over the thermal radial length scale.
	Diffusivities are calculated using the realistic solar-like diffusion profiles of \citet{brown2011}.
	(Fourth panel) An estimate of the luminosity of these thermals, normalized by the solar luminosity, is shown.
	(Fifth panel) An estimate of the filling factor of thermals required to carry the solar luminosity.
    \label{fig:thermal_sun_comparison} }
\end{figure}

Because solar downflows likely break up into thermals of various sizes, we will examine the fate of thermals with dimensionless buoyancy in the interval $[B_\ell, B_u] = 0.5B_\text{th}, 2B_\text{th}$.
Interpolating and extrapolating the data in Table \ref{table:parameters}, we use Eqns.~\ref{eqn:r_theory} \& \ref{eqn:theory_z} to calculate theoretical predictions for how thermals with these buoyancies would evolve over extended atmospheres like the solar convection zone.
Using a simple solar interior model calculated using MESA \citep{paxton&all2011} to map the density profiles of our simulation domains onto the density profile of the Sun, we plot the evolution of thermals in our estimated interval in Fig. \ref{fig:thermal_sun_comparison}.
In the first and second panels of Fig. \ref{fig:thermal_sun_comparison}, we show the evolution of these thermals' radii and velocities inside of the Sun, and compare their radial evolution to pure horizontal compression and their velocity evolution to the speed of sound.
We use the solar diffusivity models of \citet{brown2011} to estimate the timescale over which the thermal would diffuse its entropy signature ($\tau_\kappa = \chi/r_{\text{th}}^2$) and the thermal freefall timescale over its own radial length scale ($\tau_{\text{ff}} = r_{\text{th}}/w_{\text{th}}$).
In the third panel of Fig. \ref{fig:thermal_sun_comparison}, we plot the ratio of these two timescales over the thermal's evolution.

We find that our estimated solar thermal is likely in the falling regime.
These thermals experience radial compression with corresponding increases in velocity, but experience much less compression than a naive approximation based on the density stratification alone.
Interestingly, the thermal's Mach number is roughly a value of 0.1 throughout the full extent of the solar convection zone.
Furthermore, throughout the thermal's fall, we find that $\tau_\kappa \gg \tau_{\text{ff}}$, and thus we do not expect the thermal to diffuse its entropic signature.
We find a fractional diffusion rate over the thermal's transit of the solar convection zone of 
$$
\int_{R_\odot}^{0.7 R_\odot} \tau_\kappa^{-1} \frac{dr}{w_{th}} = [6.55, 1.57]\times 10^{-3}\,\text{for}\,[B_\ell, B_u].
$$
Thus, the thermal loses less than 1\% of its entropic signature to diffusion.

We now briefly estimate the enthalpy flux carried by a thermal.
Following \citet{brandenburg2016}, we estimate the enthalpy flux of the thermal as $F_{\text{enth}} = \rho_0(z_{\text{th}}) T_0(z_{\text{th}}) w_{\text{th}} S_{1,\text{th}}$.
We estimate the thermal's entropic signature as $S_{1, \text{th}} = B_{\text{th}}/(\rho_0(z_{\text{th}}) V_{\text{th}} g/c_P)$, where the thermal volume is $V_{\text{th}} = m r_{\text{th}}^3$.
We take $\rho_0(z_{\text{th}})$, $T_0(z_{\text{th}})$, and $g/c_P$ from our solar MESA model at the depth of the thermal over its evolution.
We take $r_{\text{th}}$ and $w_{\text{th}}$ to be the thermal radius and velocity shown in the first and second panels of Fig.~\ref{fig:thermal_sun_comparison}.
We find $m$ from the data in Table \ref{table:parameters} in the same manner as we found $B_{\text{th}}$ for these solar thermals.
To estimate the total luminosity carried by one thermal, we calculate $L_{\text{enth, th}} = \pi r_{\text{th}}^2 F_{\text{enth}}$, and we plot this value as a function of depth, normalized by the solar luminosity, in the fourth panel of Fig.~\ref{fig:thermal_sun_comparison}.

From this estimate of the luminosity carried by one thermal, we now calulate the filling factor of downward propagating thermals required to carry the solar luminosity.
We calculate $f_{\text{th}} = [(L_{\odot}/L_{\text{enth, th}}) (\pi r_{\text{th}}^2)]/(4\pi R^2)$, where $R$ is the thermal's radial distance from the center of the Sun.
The final panel of Fig.~\ref{fig:thermal_sun_comparison} displays $f_{\text{th}}$ vs.~depth.
The filling factor of thermals required to carry $L_\odot$ at the solar surface is greater than unity; this is unsurprising, as we know that solar surface convection carries the solar luminosity through the combined effects of upflows and downflows.
A few percentage of the solar radius beneath the photosphere, $f_{\text{th}}$ drops to a very modest $10^{-4}$, and by the base of the solar convection zone approaches $10^{-7}$.
These estimates suggest that even if a large fraction of the thermals launched from the solar surface break apart due to turbulence, entropy rain could still efficiently carry the solar luminosity deep in the convection zone.

We briefly note that our handling of the enthalpy flux here ignores the contributions of the kinetic energy flux, potential energy flux, and viscous flux.
While we expect the last of these to be inconsequential, the kinetic energy flux (which transports luminosity inwards) and the potential energy flux (which transports luminosity outwards) may be large.
We encourage future work to examine these fluxes more thoroughly, but such an examination is outside of the scope of this work.

\section{Summary \& Conclusion}
\label{sec:discussion}
In this paper we developed a simple theory of the evolution of negatively buoyant vortex rings in stratified atmospheres.
This theory predicts that dense thermals experience less entrainment than Boussinesq thermals due to increasing atmospheric density with depth.
Likewise, these thermals experience less compression than would be expected due to pure atmospheric compression of a neutrally buoyant vortex ring.
We performed 2D anelastic \& 3D fully compressible simulations of thermal evolution in the laminar regime for varying degrees of stratification, and showed that our parameterized theory describes the evolution of thermals in these systems remarkably well.
We found excellent agreement between the 2D \& 3D simulations.
The evolution of dense thermals in stratified domains is complex, but can be classified into a near-Boussinesq ``stalling'' and a high-stratification ``falling'' regime.
We estimate that solar downflows would fall into this latter regime.

The ``entropy rain'' hypothesis states that narrow downflows can transport the luminosity of the Sun via enthalpy fluxes.
If the rain stalls near the surface or its entropy diffuses away before it hits the bottom, then it cannot transport the flux.
We find that with our more accurate model for thermal propagation, solar thermals should maintain their entropy all the way to the base of the convection zone.
Hence, entropy rain is a possible mechanism for transporting the solar luminosity.

$\,$\newline$\,$
\begin{acknowledgements}
We'd like to thank Brett McKim and Nadir Jeevanjee for discussions about thermals and the laminar entrainment picture.
We'd like to further thank Nadir Jeevanjee, Axel Brandenburg, Jeremy Goodman, and the anonymous referee whose careful comments greatly improved the scientific quality and clarity of this manuscript.
This work was supported by NASA Headquarters under the NASA Earth and Space Science Fellowship Program -- Grant 80NSSC18K1199.
This work was additionally supported by NASA LWS grant NNX16AC92G and by the National Science Foundation under grant No.~1616538. 
DL is supported by Princeton Center for Theoretical Science and Lyman Spitzer Jr.~postdoctoral fellowships.
Computations were conducted with support by the NASA High End Computing (HEC) Program through the NASA  Advanced Supercomputing (NAS) Division at Ames Research Center on Pleiades with allocation GID s1647.
\end{acknowledgements}

\appendix
\section{Thermal Measurements}
\label{appendix:measurements}
Throughout this work, we frequently report the thermal's radius or its height.
We measure the thermal's radius and height as the radius from the axis of symmetry and the height above $z = 0$ at which the thermal's vortex core is located.
We assume that the vortex core is located at the thermal's entropy minima.
To find the entropy minima vertically, we integrate $\int\rho S_1 r dr$ in our Dedalus domain, then use the spectral data of that profile to sample it on a 4096-point vertical grid; we take the location of the minima on that grid to be the thermal height.
To find the entropy minima horizontally, we integrate $\int \rho S_1 dz$ in our Dedalus domain, then sample the spectral data onto a 2048-point radial grid, and take the minima of that profile to be the radius of the thermal.
For our 2D simulations, we use entropy data from the full simulation domain to perform these calculations.
For our 3D simulations, we assume that the vortex ring is azimuthally symmetric, and thus use the entropy data in the $y = 0$ plane at radial values of $x \geq 0$.
In order to find the thermal's velocity as a function of time, we use a five-point stencil to differentiate the thermal's depth, $d_{\text{th}}$,
$$
w_{\text{th}}(t) = \frac{d }{dt}d_{\text{th}}(t) = \frac{-d_{\text{th}}(t + 2\Delta t) + 8d_{\text{th}}(t + \Delta t) - 8 d_{\text{th}}(t - \Delta t) + d_{\text{th}}(t - 2\Delta t)}{12\Delta t}
$$

Calculating integral quantities such as the circulation, $\Gamma$, the buoyancy, $B$, and the volume, $\mathcal{V}$, require knowledge of what fraction of the domain constitutes the thermal.
We use the thermal tracking algorithm described in appendix \ref{appendix:tracking} to determine the radial contour, $\mathcal{C}$, that outlines the thermal as a function of height.
We then use this contour to find our integral quantities,
\begin{equation}
\Gamma = \int_0^{L_z}\int_0^{\mathcal{C}} (\grad\times\bm{u})_\phi dr\,dz, \qquad
B      = 2\pi \int_0^{L_z}\int_0^{\mathcal{C}} \rho S_1 r dr\,dz, \qquad
\mathcal{V} = 2\pi \int_0^{L_z}\int_0^{\mathcal{C}}r dr\,dz.
\end{equation}

\section{Thermal Tracking Algorithm}
\label{appendix:tracking}
We use a thermal tracking algorithm very similar to the one used in  \citet{lecoanet&jeevanjee2018} and inspired by the work of \citet{romps&all2015} in order to determine the full extent of the thermal, as pictured by the elliptical outlines in Fig.~\ref{fig:evolution_colormeshes}. 
We begin by measuring the thermal's velocity versus time, $w_{\text{th}}$, as described in appendix \ref{appendix:measurements}. 
We calculate the streamfunction of the velocity field as in \citet{romps&all2015},
\begin{equation}
\frac{\partial \psi}{\partial r} = 2\pi \rho r (w - w_{\text{th}}),
\label{eqn:thermal_tracker}
\end{equation}
using vertical velocity data, $w$, in a 2D domain which radially spans $r = [0, L_r]$ and vertically spans the same depth as the simulation domain.
For our 2D simulations, this is simply the output of the simulations.
For our 3D simulations, we assume that the vortex ring is azimuthally symmetric, and thus use the vertical velocity data in the $y = 0$ plane at values of $x \geq 0$.
We solve Eqn.~\ref{eqn:thermal_tracker} with the boundary condition that $\psi = 0$ at $r = 0$. 
The contour $\psi = 0$ is taken to be the contour bounding the thermal, $\mathcal{C}$.

\section{Table of Simulations}
\label{appendix:table}
Information regarding the simulation resolution, evolution time, and CFL safety factor for each of the simulations presented in this work is contained in table \ref{table:simulation_info}.
The Python scripts used to perform all simulations and analysis in this work are stored online in a Zenodo repository \citep{supp_andersetall2019} at \dataset[10.5281/zenodo.3311894]{https://doi.org/10.5281/zenodo.3311894}.
The 3D, $N_\rho = 3$ simulation displays spectral instabilities at very late times as the simulation is under-resolved at these times.
This affects thermal radial measurements at depths $\geq 17.5$.
We therefore truncate the data from that simulation at a depth of 15, which corresponds to multiple freefall times before these instabilities affect the solution.

\begin{deluxetable*}{c c c c c }
\tabletypesize{\footnotesize}
\caption{Table of simulation information
\label{table:simulation_info}
}
\tablehead{																																															
\colhead{$N_\rho$} & \colhead{nr or nx = ny} & \colhead{nz} & \colhead{$t_{\text{evolution}}$} & \colhead{safety}	}	
\startdata																																															
\multicolumn{5}{l}{2D Anelastic Simulations}\\
0.1 	& 		128			& 512			& 50 	&	0.6	\\
0.5 	& 		128			& 512			& 45 	&	0.6	\\
1	 	& 		128			& 512			& 41 	&	0.6	\\
2	 	& 		128			& 1024			& 34	&	0.4	\\
3	 	& 		192			& 1024			& 29	&	0.4	\\
4	 	& 		256			& 1024			& 26 	&	0.3	\\
5	 	& 		256			& 1536			& 25 	&	0.14	\\
6	 	& 		256			& 1536			& 23 	&	0.08	\\
\multicolumn{5}{l}{3D Fully Compressible Simulations}\\
0.1 	& 		256			& 512			& 50 		&	0.15	\\
0.5 	& 		256			& 512			& 45 		&	0.15	\\
1	 	& 		256			& 512			& 41	 	&	0.15	\\
2	 	& 		384			& 1024			& 34    	&	0.1	\\
3	 	& 		384			& 2048			& 25    	&	0.15	\\
\enddata																																															
\tablecomments{
Our 2D cylindrical sims have spectral mode resolution of nr x nz, and our 3D cartesian sims have spectral mode resolution of nx x ny x nz.
We evolve our simulations from initial conditions until $t_{\text{evolution}}$ dimensionless freefall times have passed.
We use a CFL to determine the appropriate size of our timestep at each timestep in our simulation, and the CFL safety factor decreases with increasing stratification.
Our 2D sims use the RK443 timestepper and our 3D sims use the SBDF2 timestepper, resulting in different safety factor sizes in 2D and 3D.
}
\end{deluxetable*}

\end{document}